\documentclass[prl,twocolumn,superscriptaddress,floatfix,showpacs]{revtex4-1}

\usepackage[utf8]{inputenc}
\usepackage[T1]{fontenc}
\usepackage{amsfonts,amsmath,amssymb,mathbbol,dsfont}
\usepackage{latexsym,times}
\usepackage{graphicx,psfrag,xcolor}
\usepackage{bm,bbm,dcolumn}
\begin{document}

\title{Exact Solution of Quadratic Fermionic Hamiltonians for Arbitrary Boundary Conditions}

\author{Abhijeet Alase}
\affiliation{\mbox{Department of Physics and Astronomy, Dartmouth
College, 6127 Wilder Laboratory, Hanover, New Hampshire 03755, USA}}

\author{Emilio Cobanera}
\affiliation{\mbox{Department of Physics and Astronomy, Dartmouth
College, 6127 Wilder Laboratory, Hanover, New Hampshire 03755, USA}}

\author{Gerardo Ortiz}
\affiliation{\mbox{Department of Physics, Indiana University,
Bloomington, Indiana 47405, USA}}

\author{Lorenza Viola}
\affiliation{\mbox{Department of Physics and Astronomy, Dartmouth
College, 6127 Wilder Laboratory, Hanover, New Hampshire 03755, USA}}

\begin{abstract}
We present a procedure for exactly diagonalizing finite-range quadratic fermionic 
Hamiltonians with arbitrary boundary conditions in one of $D$ dimensions, 
and periodic in the remaining $D-1$. The key is a Hamiltonian-dependent
separation of the bulk from the boundary. By combining information from the 
two, we identify a matrix function that fully characterizes the solutions, 
and may be used to construct an efficiently computable indicator of bulk-boundary correspondence. 
As an illustration, we show how our approach correctly describes the zero-energy Majorana
modes of a time-reversal-invariant $s$-wave two-band superconductor in a Josephson ring configuration, 
and predicts that a fractional 4$\pi$-periodic Josephson effect can only be observed in phases hosting 
an odd number of Majorana pairs per boundary.
\end{abstract}

\date{\today}
\maketitle

Developing a quantitative understanding of the physical 
properties of fermionic systems in the presence of non-trivial boundaries 
has widespread significance from both a fundamental and applied 
perspective. 
Not only has the behavior of fermions at a boundary 
informed leading material-characterization techniques like
angle-resolved photoemission spectroscopy \cite{arpes} and 
the revolution in metrology brought about by 
the integer quantum Hall effect \cite{Klitzing}; nowadays,
surface states of topological insulators 
and Majorana boundary modes of topological superconductors 
\cite{kane,Bernevig} play a central role in state-of-the-art 
proposals ranging from coherent spintronics \cite{topospintronics,Nitin} 
to topological quantum computation \cite{Kitaev,pachos}.
  
All of the above phenomena  
are linked by a common theme: topologically non-trivial band structures 
\cite{Bernevig}. Band structure theory, including the topological classification of 
mean-field fermionic systems \cite{kitaev}, rests on a 
manifestation of crystal translational symmetry, the Bloch theorem. Since translational 
symmetry is broken by the presence of a boundary, it is remarkable 
that there exists a connection between the topological nature of the bulk 
and the boundary physics -- the 
{\em bulk-boundary (BB) correspondence} \cite{Bernevig,ryuLetter}. 
This principle states that a topologically non-trivial 
bulk mandates the emergence of fermionic states localized on the boundary,
when boundary conditions (BCs) are changed from periodic to open, and that 
such states are distinguished by their {\em robustness} against symmetry-preserving
local perturbations.  
While this heuristics has been numerically validated in a variety of cases, and rigorous 
results exist for discrete-time systems described by one-dimensional quantum 
walks \cite{Werner}, no  general analytic insight 
is available as yet.  Allowing for {\em arbitrary} BCs 
is necessary for any theory of BB correspondence to 
capture the robustness of the emerging localized modes to different perturbations
\cite{ours2}. Further motivation stems from studies of quantum quenches \cite{hegde14,fagotti}, 
where robustness against changes of the BCs has been 
argued to control the (quasi)local symmetries that characterize the stationary properties 
in the bulk.  Tackling these issues calls for a procedure to determine energy eigenstates of lattice 
Hamiltonians with arbitrary BCs, comparable in conceptual and computational power to 
what the Fourier transform accomplishes in the periodic case.

In this work, we introduce a methodology for diagonalizing 
in closed form finite-range quadratic fermionic Hamiltonians with translational 
symmetry broken by arbitrary BCs.  Our central insight is  
a generalization of  Bloch's theorem built on the recognition that a useful 
separation of the bulk from the boundary should be model-dependent. 
We identify an {\em indicator for BB correspondence},  
that exploits both information about the bulk -- encoded
in ``generalized Bloch states'' -- and the nature of the boundary -- 
encoded in a ``boundary matrix''.
For periodic BCs, we prove that, generically, 
such indicator predicts {\em no} localized
edge state irrespective of the bulk structure.  As an application,  
we explore the Josephson response of a
$s$-wave, time-reversal-invariant two-band topological superconductor (TS)
introduced in \cite{Viola}, 
and show how the boundary matrix reveals that a fractional 
Josephson effect occurs
{\em only} in the phase with one pair of Majoranas per boundary, consistent with 
the physical picture based on fermion parity switches \cite{ours1}. 
Mathematically, our approach generalizes existing algorithms for diagonalizing 
banded Toeplitz matrices \cite{Trench} to the block-Toeplitz case with arbitrary corner 
modifications, with complexity independent upon system size.

\emph{Model Hamiltonians.--} Consider fermionic systems defined on a one-dimensional lattice 
consisting of $j=0,\ldots, L-1$ identical cells, each containing $m=1, \ldots,d$ internal degrees of 
freedom, associated for instance to spin and orbital motion.
Let the creation (annihilation) operator for mode labeled by $(j,m)$ 
be denoted by $c_{j,m}^{\dagger}$ ($c_{j,m}$), and let 
$\psi_{j}^{\dagger} \equiv [ c_{j,1}^{\dagger} \: c_{j,2}^{\dagger} \: \cdots \: c_{j,d}^{\dagger} 
\: c^{\;}_{j,1} \: c^{\;}_{j,2} \: \cdots \; c^{\;}_{j,d}]$ be the corresponding $(2d)$-dimensional 
Nambu vector. We consider finite-range $R$, $R \ll L$, disorder-free quadratic Hamiltonians of the form
\begin{equation}
\widehat{H}= \frac{1}{2} 
\sum_{r=0}^{R}\Big(\sum_{j=0}^{L-r-1}\psi_{j}^{\dagger}h_{r}\psi^{\;}_{j+r}
+\sum_{j=L-r}^{L-1}\psi_{j}^{\dagger}g_{r}\psi^{\;}_{j+r-L}+ \text{h.c.}\Big),
\label{many-body Hamiltonian}
\end{equation}
where the matrices $h_{r}$ and $g_r$ describe hopping and pairing 
among fermions situated $r$ cells apart in the bulk and, respectively, at the boundary.
In this way, standard periodic and open BCs
correspond to $g_{r}=h_{r}$ and $g_{r}=0$, $ \forall r$. 
Hamiltonians of the form \eqref{many-body Hamiltonian} arise ubiquitously 
in mean-field descriptions of fermionic systems as realized in both solid-state and cold-atom 
platforms \cite{Ripka,DFT,Liang}.

Analyzing the single-particle sector of $\widehat{H}$ suffices 
to study its many-body spectrum \cite{Ripka}. That is, we let 
$\widehat{H} =  { \frac{1}{2} } \Psi^\dagger H \Psi$, with 
$\Psi^\dagger \equiv [\psi^\dagger_0\: \ldots \: \psi^\dagger_{L-1}]$.  In this way,  
the Hilbert space ${\mathcal H}$ on which the single-particle Hamiltonian ${H}$ acts 
may be conveniently factorized into the tensor product of two subsystems, 
${\mathcal H} \simeq {\mathbb C}^L \otimes {\mathbb C}^{2d}\equiv {\mathcal H}_L 
\otimes {\mathcal H}_I$, associated to lattice and internal factors.
Let the operators $c_{j,m}$ and $c_{j,m}^{\dagger}$ be associated with 
vectors $\left|j\right\rangle \left|m\right\rangle$
and $\left|j\right\rangle \left|m+d\right\rangle$, respectively.
In the basis $\left\{ \left|j\right\rangle \left|m\right\rangle |\ 0\le j\le L-1,\ 1\le m\le2d\right\} $,
${H}$ is given by 
\begin{equation}
{H}=\sum_{r=0}^{R}\left(T^{r}\otimes h_{r}+
\left(T^{\dagger}\right)^{L-r}\otimes g_{r}+ \text{h.c.}\right),
\label{BdG Hamiltonian}
\end{equation}
where $T$ is the left-shift operator 
$T\left|j\right\rangle  \equiv \left|j-1\right\rangle$,  
$\forall j\ne0$, $T\left|0\right\rangle \equiv 0$, and 
$T^{\dagger}$ implements the corresponding right shift.
Thus, ${H}$ is a ``corner-perturbed'' banded block-Toeplitz
matrix with $2R+1$ bands.  Namely, the $r$th off-diagonal bands above and
below the diagonal have blocks given by bulk interaction matrices
$h_{r}$ and $h_{r}^{\dagger}$, respectively, whereas boundary 
terms appear in the corner of the matrix. The $\left(L \! -\! r\right)$th
off-diagonal bands above and below the diagonal, which lie close to
the corners, consist of blocks given by $g_{r}^{\dagger}$ and $g_{r}$, 
respectively. 

\emph{Periodic boundary conditions revisited.--} 
Periodic BCs are employed in calculations of band structure and 
bulk topological invariants alike \cite{Bernevig}. In this case, the single-particle 
Hamiltonian ${H}$ in Eq. \eqref{BdG Hamiltonian} is a circulant block-Toeplitz matrix, 
which may be expressed as 
$H = \sum_{r=0}^{R} \, (V^{r}\otimes h_{r}+ \text{h.c.})$, in terms of 
the cyclic left-shift operator $V \equiv T+ (T^{\dagger} )^{L-1}$.  Crucially, 
translational symmetry implies 
that $H$, $V$, and $V^{\dagger}$ form a commutative set,
allowing for the eigenspectrum of $H$ to be determined via standard discrete 
Fourier transform from the lattice to the momentum basis on ${\mathcal H}_L$. 
For later reference, let us
introduce the generalized {\em $z$-transformed lattice basis}, 
\begin{equation}
\left |z \right\rangle  \equiv \frac{1}{\sqrt{N(z)}}\sum_{j=0}^{L-1}  z^{j} | j \rangle ,
\quad z \in \mathbb{C}, z \ne 0, 
\label{Generalized eigenvector}
\end{equation}
where $N(z)$ is a normalization constant, and define the 
``reduced bulk Hamiltonian'' $h_B(z)$ as the matrix-valued symbol
\cite{Symbol} of the block-Toeplitz matrix without boundary terms:  
\begin{equation}
h_B(z) \equiv \sum_{r=0}^{R} \, (z^{r}h^{\;}_{r}+z^{-r}h_{r}^{\dagger}).
\label{bulk Hamiltonian}
\end{equation}
The generalized discrete Fourier transform in Eq. \eqref{Generalized eigenvector} 
associates $z$ to the pseudo-momentum $k$, with $z \equiv e^{ik}$ and 
$k \equiv 2\pi q/L,\ q\in \{ 0,1,\ldots,L-1 \}$, defining the (first) Brillouin zone. 
Then, the eigenvectors of $H$ may be expressed as
$|\epsilon \rangle \equiv  |z \rangle |u (\epsilon,z)\rangle$, 
where  $\left|u\left(\epsilon,z\right)\right\rangle$
is the eigenvector of the reduced bulk Hamiltonian $h_B (z)$
with eigenvalue $\epsilon$ -- which is simply a reformulation of the familiar Bloch theorem. 
The cyclic shift symmetry restricts solutions to the Brillouin zone, 
and $z$ to lie on the unit circle.
Therefore, by diagonalizing $h_B(z)$ for all $q$, 
the complete quasi-particle energy spectrum and the corresponding 
eigenvectors may be obtained.

As lattice translation ceases to be a symmetry, the discrete Fourier
transform fails to diagonalize ${H}$. In particular, the left and  
right shift operators $T$ and $T^{\dagger}$ {\em do not} share a common eigenbasis,
calling for a different diagonalization approach. We next introduce a new diagonalization 
method that relies on a mapping of the Brillouin zone to the full complex plane. 

\emph{Bulk-boundary separation and bulk equation.--}  Hamiltonians with arbitrary
BCs are locally symmetric under left and right shifts in the bulk, however, these symmetries 
are explicitly broken at and near the boundaries. The crux of our approach consists of 
separating bulk from boundary subsystems. To this end, we define orthogonal projectors onto 
the bulk, 
${\mathcal P}_{B}\equiv \sum_{j=R}^{L-R-1} |j \rangle \langle j |$, 
and onto the boundary, 
${\mathcal P}_{\partial} \equiv \mathbb{1}_L-{\mathcal P}_{B}$, 
where $\mathbb{1}_L$ is the $L$-dimensional identity operator on ${\mathcal H}_L$. 
The eigenvalue equation for ${H}$ then splits into a bulk and a boundary equation: 
${\mathcal P}_{B}{H} |\epsilon \rangle =
\epsilon{\mathcal P}_{B} |\epsilon \rangle$, and 
${\mathcal P}_{\partial} H |\epsilon \rangle =
\epsilon{\mathcal P}_{\partial} |\epsilon \rangle$. 
The advantage of such a separation is that one obtains 
{\em simultaneous} (relative) eigenvectors of the bulk-projected $T$ and $T^{\dagger}$ 
operators. The resulting eigenvalue equations are: 
${\mathcal P}_{B}T^{r} |z \rangle = z^{r}{\mathcal P}_{B} |z \rangle$, 
${\mathcal P}_{B}(T^{\dagger})^{r'} |z \rangle = z^{-r'}{\mathcal P}_{B}  |z \rangle$, 
$\forall r,r'\le R$, while 
${\mathcal P}_{B}T^{r} =0= {\mathcal P}_{B}(T^{\dagger})^{r'}$, 
$\forall r,r'\ge L-R$.
As for periodic BCs, it follows that these ``generalized Bloch states'' are of product form 
$|z \rangle |u (\epsilon,z) \rangle $,
where as above $|u (\epsilon,z) \rangle $ is an
eigenvector of $h_B (z)$ with eigenvalue $\epsilon$. 

By construction, 
$h_B (z )$ is a {\em small} matrix, of dimension $2d \times 2d$. 
If ${\mathbb 1}_I$ denotes the $2d$-dimensional identity operator on ${\mathcal H}_I$, 
the relevant characteristic equation establishes a functional relationship between $\epsilon$ and $z$, 
of the form
\begin{equation} 
P(\epsilon,z) \equiv z^{2dR}\det\left[\left(h_B (z)-\epsilon {\mathbb 1}_I \right)\right]=0,
\label{Dispersion relation}
\end{equation}
where the prefactor $z^{2dR}$ ensures that $P(\epsilon,z)$ is a bi-variate polynomial in 
$\epsilon$ and $z$, of degree at most $(2R)(2d)=4dR$.
In general, there may exist multiple generalized Bloch states corresponding to a given value
of $\epsilon$. 
Let $z_{\ell}(\epsilon)$, $\ell=1,\ldots,n$, denote the (non-zero) distinct roots of Eq. 
\eqref{Dispersion relation} for the given $\epsilon$, and $s_\ell(\epsilon)$ the corresponding number of linearly 
independent eigenvectors of $h_B(z_\ell)$ (that is, the nullity 
of $(h_B (z_\ell)-\epsilon {\mathbb 1}_I$)).
The eigenvectors of ${H}$ may then be written as linear combinations of 
degenerate generalized Bloch states:
\begin{equation}
|\epsilon \rangle \equiv \sum_{\ell=1}^n \sum_{s = 1}^{s_\ell} \alpha_{\ell, s}\, 
| z_\ell (\epsilon) \rangle |u_s (\epsilon,z_\ell ) \rangle ,
\quad\alpha_{\ell,s} \in \mathbb{C} .
\label{Ansatz}
\end{equation}
In this way, the solutions of the bulk equation provide an Ansatz for
the eigenvectors of $H$, where the amplitudes $\alpha_{\ell,s}$
are yet to be determined \cite{next}. 
Of particular interest is the Ansatz for $\epsilon=0$, which provides a 
family of {\em possible} zero-energy modes of the system, independent of the BCs.
Notice that the solution of the bulk equation alone does {\em not} imply 
the existence of an excitation at a given value of $\epsilon$, unless the 
boundary equation is simultaneously satisfied.

In general,  $h_B(z)$ is {\em not} Hermitian, except on the unit circle $|z|=1$. 
This is not surprising since such an effective Hamiltonian represents an open system, 
with no boundaries and no torus topology  \cite{Non-hermiticity}. 
Generalized Bloch states exist for every $z \ne 0$, realizing 
an over-complete set of solutions of the bulk equation. Those consistent with
values of $z$ on the unit circle are in one-to-one correspondence
with the solutions of the infinite periodic system. The rest correspond
to solutions with exponential behavior,
providing a continuation off the Brillouin zone. 
Thus, Eq. \eqref{Ansatz} may be regarded as a generalization of the Bloch 
theorem to arbitrary BCs, with Eq. \eqref{Dispersion relation} providing a natural 
analytic continuation of the dispersion relation. 

\emph{Boundary equation and emergence of localized modes.--} 
The Ansatz \eqref{Ansatz} yields an eigenvector of ${H}$ 
only if the boundary equation, 
${\mathcal P}_{\partial}\left(H - \epsilon\mathbb{1}\right) |\epsilon\rangle =0$, 
is satisfied for appropriate $\alpha_{\ell,s}$, with $\mathbb{1}$ denoting the identity operator 
on ${\mathcal H}$.
For notational simplicity, let us assume that $s_\ell =1,$ $\forall \ell, \epsilon$, 
so that $\alpha_{\ell,s}\equiv \alpha_\ell$ (see \cite{note,next} 
for the general case).
To make the action of the projector ${\mathcal P}_{\partial}$
explicit, it is convenient to isolate the $4dR$ basis states in ${\mathcal H}$ 
that correspond to lattice sites on the boundary, by letting 
$\{ |b\rangle  \equiv |j\rangle |m\rangle \,|\, 0 \le j\le R-1, L-R \le j\le L-1; 1\le m \le 2d \}$.
The above boundary equation may then be rewritten as 
$\sum_{\ell =1}^n \alpha_{\ell} B_{L, b \ell}(\epsilon)=0$, where the 
``boundary matrix'' $B_{L}(\epsilon)$ for size $L$ and energy $\epsilon$ is 
the (generally non-square) $4dR \times n$ matrix with entries given by
\begin{equation}
\label{boundary Hamiltonian}
\hspace*{-1mm}
[B_L(\epsilon)]_{b \ell} = B_{L, b\ell} (\epsilon)  \equiv 
\langle b| (H-\epsilon\mathbb{1}) |z_\ell (\epsilon) \rangle| u_1(\epsilon,z_{\ell})\rangle.
\end{equation}
By construction, any set of values of $\{\alpha_\ell\}$ satisfying Eq. (\ref{boundary Hamiltonian})
is a vector in the kernel of $B_L(\epsilon)$. Thus, the boundary equation may be 
restated as $\det \, [ B^\dagger_L(\epsilon)  B_L (\epsilon) ]=0$. 
When this condition is obeyed, $\epsilon$ is an eigenvalue of $H$,  
with degeneracy equal to the nullity of $B_L(\epsilon)$. 

For fixed BCs (fixed $g_{r}$), the localized modes of the system and 
their energies show asymptotic behavior in the thermodynamic limit, 
$L\rightarrow\infty$, which our method enables to characterize analytically.
A localized mode in the thermodynamic limit has constituent generalized Bloch 
states with $\left|z_{\ell}\right|\ne1$. This allows a simplification in 
the boundary matrix, as any $L$-dependent terms in $B_L(\epsilon)$
may be replaced by the appropriate limit, given by $\lim_{L\rightarrow\infty}z_{\ell}^{L}\rightarrow0$
and $\lim_{L\rightarrow\infty}z_{\ell}^{-L}\rightarrow0$ for $\left|z_{\ell}\right|<1$
and $\left|z_{\ell}\right|>1$, respectively. The existence check for
localized modes and their calculations is then carried out in the same
way as in the finite-$L$ case.

\emph{An indicator of bulk-boundary correspondence.-- } 
The boundary matrix defined in Eq. \eqref{boundary Hamiltonian} points to a natural 
strategy for constructing useful indicators of BB correspondence based on  
combined information from both the bulk and the boundary. In particular,  
for zero-energy modes, we propose 
\begin{equation}
\mathcal{D} \equiv \log \left(\det [ B^\dagger_{\infty} (0) B_\infty (0) ] \right) 
\label{bulk-boundary indicator}
\end{equation}
as one such indicator for an infinite system. We claim that the existence of 
zero-energy edge modes manifests as a singularity in the value of $\mathcal{D}$. 
Consistent with this claim, we can rigorously prove that, under generic assumptions 
on the matrix $h_{r}$ for $r=R$,
the indicator ${\mathcal D}$ is {\em always finite} 
under periodic BCs, irrespective of the bulk properties, see \cite{note}. 
We remark that other indicators are also in principle applicable to systems where 
translational invariance is broken (notably, based on Pfaffians 
\cite{Kitaev01,beenakker11}); even for clean systems as we consider, however, 
their numerical evaluation becomes 
computationally demanding for large system size. 

\emph{Example.--} As a first illustration,
we revisit the paradigmatic case of Kitaev's $p$-wave TS chain with open BCs 
\cite{Kitaev01} (see \cite{note} for full detail). 
With reference to Eq. (\ref{BdG Hamiltonian}), this nearest-neighbor model corresponds to $R=1$ and 
$2 \times 2$ matrices $2 h_0=-\mu\sigma_z$, $h_1=-t\sigma_z + i\Delta\sigma_y$, $g_1\equiv 0$, 
where $\mu, t, \Delta$ denote chemical potential, hopping, and superconducting pairing respectively, and 
$\sigma_\nu,\nu=x,y,z$, are  Pauli matrices in the Nambu basis. For any $\epsilon$, the characteristic 
equation for $h_B(z)$, Eq. \eqref{Dispersion relation}, is quartic in $z$, 
which enables a closed-form solution by radicals.
The roots appear as two reciprocal pairs, say, $\left\{ z_{1},z_{1}^{-1},z_{2},z_{2}^{-1}\right\} $, 
where $\left|z_{1}\right|, |z_2| \le1$. For finite chain length $L$, the boundary equation is satisfied if any 
of the two equalities $f_{\pm}(z_1)=f_{\pm}(z_2)$ hold, where $f_{\pm}(z)=[{b(z)}/( {\epsilon +a(z)} )]\,
[ ({ 1+z^{L+1} }) /( { 1-z^{L+1} }) ]^{\pm1}$, and $a(z)=\mu+t (z+z^{-1}),b(z)=\Delta (z-z^{-1})$ \cite{Remark2}.
In the thermodynamic limit, the condition for zero-energy edge modes simplifies to 
$a(z_{1})/b(z_{1})=a(z_{2})/b(z_{2})$, which is satisfied if and only if $|\mu|< |2t |$.
This parameter regime, with phase boundary $|\mu|=|2t |$, defines the 
topologically non-trivial phase, hosting one Majorana mode per edge.
If $\mu^{2} < 4| t^{2}-\Delta^{2}|$, $(z_{1},z_{2})$ form a complex conjugate pair, 
whereas for $\mu^{2} > 4| t^{2}-\Delta^{2} |$, both $z_{1}$ and $z_{2}$ are real. 
The self-adjoint Majorana mode localized on the left edge is then given by 
$\gamma \equiv \sum_{j=1}^{\infty} (z_{1}^{j}-z_{2}^{j})(c_{j}\mp c_{j}^{\dagger}),$ 
for $\mu^{2}   \lessgtr 4| t^{2}-\Delta^{2} |$.

\emph{Josephson effect in two-band $s$-wave superconductors.--}
The Kitaev chain in its topologically non-trivial phase is known to exhibit a fractional 
Josephson effect \cite{Kitaev01}, that is, the Josephson current is $4 \pi $-periodic 
(more generally, $2\pi l$-periodic, with integer $l>1$) as a function of the superconducting 
phase difference $\phi$, and the many-body energy $E(\phi)$ correspondingly switches 
parity \cite{ours1}. Such an effect is regarded as both a hallmark and a leading 
observable signature of topological superconductivity.   The simplicity of the 
topological phase diagram in the Kitaev chain (either 0 or 1 Majorana mode per edge) 
allows for an unambiguous association between a trivial (non-trivial) phase and a standard 
(unconventional) Josephson response; however, it is not {\em a priori} obvious what to expect for more 
complex TSs, which may support phases with different numbers of Majorana modes 
-- or, respectively, different numbers of Majorana {\em pairs} per edge, if time-reversal 
symmetry is preserved \cite{Bernevig}. 
As we show next, the existence of localized Majorana modes  
does {\em not} suffice, in general, for the system to display fractional Josephson effect. 

Consider the time-reversal-invariant two-band $s$-wave TS wire introduced in 
\cite{Viola}. Based on both the original numerical solution under open BCs and 
analysis of the appropriate boundary matrix $B_L(0)$ \cite{note}, phases with 
zero, one, or two pairs of (helical) Majorana modes localized on each boundary 
may exist.  Using a partial Berry-phase parity as a topological indicator \cite{Viola}, 
only the phase hosting one pair of Majorana modes is predicted to be 
topologically non-trivial.  Thanks to the present analytic approach, in particular the 
BB indicator ${\mathcal D}$ defined in Eq. \eqref{bulk-boundary indicator}, 
we are now in a position to correlate this prediction with the more physical 
-- in principle experimentally accessible -- Josephson response of the system. 

\begin{figure}[t]
\includegraphics[width=8cm,height=8cm]{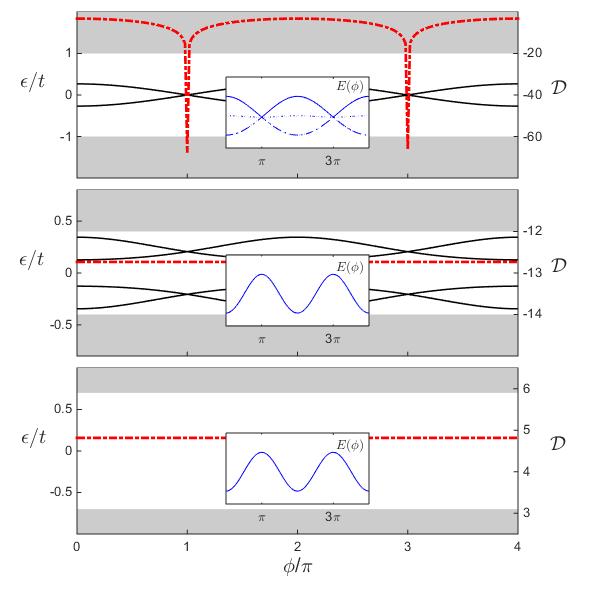} 
\vspace*{-5mm}
\caption{ \label{fig:1} 
(Color online) Quasi-particle energies
(solid black lines) and BB indicator ${\mathcal D}$ [Eq. \eqref{bulk-boundary indicator}]
(dashed red line) vs. flux $\phi$ in a phase supporting one (top), two (middle), 
and zero (bottom) pairs of Majorana modes per edge. The grey shaded regions 
indicate bulk quasi-particle energies.  Insets: Many-body ground state energy $E(\phi)$.
With reference to \cite{note}, the parameter 
values are: $t=\lambda=\Delta=1$, $\mu=0$, $w=0.2$, $u_{cd}=2$ (top), $u_{cd}=0.6$ (middle), 
$u_{cd}=3.7$ (bottom). 
%For both the single-particle and the many-body energy, 
In all calculations, lattice size $L= 60$ is used.
}
\end{figure}

Our results are summarized in Fig. \ref{fig:1}.  As the insets show, 
a fractional Josephson effect emerges {\em only} in the phase that is predicted to 
be topologically non-trivial according to its partial Berry-phase (odd) parity \cite{note}.
Some physical insight may be gained by looking at the dependence of quasi-particle 
energy upon flux $\phi$: as seen in the top panel, 
the $4\pi$-periodicity is associated with a crossing of a positive and a 
negative quasi-particle energy level; this crossing occurs precisely at zero energy,
indicating the presence of a pair of Majorana modes for the value of flux at the crossing 
(solid black lines).  In contrast, in the trivial phase with two pairs of Majorana modes 
(middle panel), the level crossing does {\em not} occur at zero energy, leading to
the standard $2\pi$-periodicity of $E(\phi)$, also found when no Majorana mode is 
present (bottom panel). Since these quasi-particle energy levels lie in the gap, they
are localized, and we can carry out the analysis in the thermodynamic
limit \cite{note}. This reveals that, in the non-trivial phase, the boundary
equation is satisfied at $\phi=\pi, 3\pi$, confirming 
the presence of exact zero energy modes at those values. 
Even more interestingly, the proposed indicator ${\mathcal D}$ has been 
numerically evaluated and plotted (dotted red lines): 
singularities clearly emerge {\em only} in the topologically non-trivial phase, 
as claimed. 

\emph{Discussion.--}  
We investigated finite-range quadratic fermionic Hamiltonians
for which translational symmetry is broken only by arbitrary BCs, and 
showed how a Hamiltonian-dependent BB separation 
can make the property of the system being ``almost translationally symmetric" 
quantitatively useful -- leading to a natural generalization of the Bloch 
theorem. Building on this, we described an efficient diagonalization algorithm 
which, for $D=1$, reduces the problem of determining the full set of 
eigenvalues and eigenvectors to one of finding the roots of 
the boundary equation.
Our algorithm successfully identifies the interplay between bulk properties
-- captured by the generalized Bloch states $|z \rangle |u (\epsilon,z) \rangle $ -- 
and BCs -- captured by the boundary matrix, $B_L(\epsilon)$, of fixed dimension 
independent of $L$.  
Since the calculation of $L$-dependent terms in $B_L(\epsilon)$ (to fixed accuracy) 
may be effected in a single computation, the complexity of our algorithm is ${\mathcal O}(1)$, 
in contrast to ${\mathcal O}[(d L)^3 ]$ for generic methods of evaluating the characteristic
polynomial of $H$.

The advantages of our algorithm extend straightforwardly to $D$-dimensional 
quadratic Hamiltonians, $D>1$, if the standard procedure of imposing periodic BCs 
in $D-1$ directions is employed: in this way, 
the model reduces to a poly-sized set of $D=1$ lattices subject 
to arbitrary BCs, to which our algorithm applies. We thus expect our 
approach to both have immediate relevance to electronic structure calculations for 
lattice systems, and to further elucidate gapless topological superconductivity 
-- notably, the emergence of Majorana flat bands and anomalous BB correspondence
uncovered in \cite{Flatband}.
For more general BCs, e.g., two open directions, our model-dependent procedure for 
BB separation and generalized Bloch theorem go through with minor 
modifications.  On the one hand, this prompts the question of whether the 
concept of a Wannier function may also be generalized for arbitrary BCs.  
On the other hand, the procedure for incorporating BCs is more involved, 
calling for separate investigation. 

Beyond equilibrium scenarios, our approach should prove 
advantageous to evaluate in closed form the unitary propagator 
$\exp(-i \widehat{H}t)$ describing free evolution under arbitrary BCs, and to 
diagonalize the Floquet propagator describing periodically driven 
fermionic systems \cite{Floquet}.  Since our algorithm does 
not exploit the Hermiticity of the Hamiltonian, a further direction of 
investigation is the application to open Fermi 
systems obeying quadratic Lindblad master equations 
\cite{prosen}, with the potential to shed light onto BB correspondence 
in engineered topological phases far from equilibrium \cite{ZollerDiss}. 
Lastly, despite important differences at the 
single-particle level \cite{Ripka}, 
our algorithm applies to arbitrary 
BCs in quadratic bosonic systems. 
This is an intriguing observation, since there is no 
BB correspondence for bosons, yet a topological 
classification might be possible in terms of generalized, symplectic 
Berry phases.

It is a pleasure to thank Alex Barnett for useful discussions and input on 
computational complexity. Work at Dartmouth was supported by the NSF through 
Grant No. PHY-1066293 and the Constance and Walter Burke Special Projects Fund 
in Quantum Information Science.

{}

\vfill

\begin{thebibliography}{100}

\bibitem{arpes} A. Damascelli, Z. Hussain, and Z.-X. Shen, 
Rev. Mod. Phys. {\bf 75}, 473 (2003).

\bibitem{Klitzing} J. Weis and K. von Klitzing,
%Metrology and microscopic picture of the integer QHE 
Phil. Trans. R. Soc. A {\bf 369}, 3954 (2011).

\bibitem{kane}
%Topological insulators
M. Z. Hasan and C. L. Kane, Rev. Mod. Phys. {\bf 82}, 3045 (2010);
%Topological insulators and superconductors
X.-L. Qi and S.-C. Zhang, ibid. {\bf 83}, 1057 (2011); 
J. Alicea, Rep. Prog. Phys. {\bf 75}, 076501 (2012); 
%Search for Majorana fermions in superconductors
C. W. J. Beenakker, Annu. Rev. Con. Mat. Phys. {\bf 4}, 113 (2013). 

\bibitem{Bernevig}
A. B. Bernevig (with T. L. Hughes), 
\textit{Topological Insulators and Topological Superconductors} 
(Princeton University Press, 2013).

\bibitem{topospintronics}
%Spintronics and pseudospintronics in graphene and topological insulators
D. Pesin and A. H. MacDonald, Nature Materials {\bf 11}, 409 (2012). 

\bibitem{Nitin} A. R. Mellnik, J. S. Lee, A. Richardella, J. L. Grab, P. J. Mintun, M. H. Fischer, 
A. Vaezi, A. Manchon, E.-A. Kim, N. Samarth, and D. C. Ralph, Nature {\bf 511}, 449 (2014).

\bibitem{Kitaev} 
A. Yu. Kitaev, Ann. Phys. {\bf 303}, 2 (2003). 

\bibitem{pachos} 
J. K. Pachos,
{\em Introduction to Topological Quantum Computation} (Cambridge, UK, 2012).

\bibitem{kitaev}
%Periodic table for topological insulators and superconductors
A. Kitaev, AIP Conf. Proc. {\bf 1134}, 22 (2009); 
%Topological insulators and superconductors: ten-fold way and dimensional hierarchy
S. Ryu, A. Schnyder, A. Furusaki, and A. Ludwig, New J. Phys. {\bf 12}, 065010 (2010).

\bibitem{ryuLetter}
%For a lucid example of how the BB correspondence
% is often justified in the literature, see
%Topological Origin of Zero-Energy Edge States in Particle-Hole Symmetric Systems
S. Ryu and Y. Hatsugai, Phys. Rev. Lett. {\bf 89}, 077002 (2002). 

\bibitem{Werner}
C. Cedzich,  F. A. Grünbaum, C. Stahl, L. Velázquez, A. H. Werner, and R. F. Werner,
%Bulk-edge correspondence of one-dimensional quantum walks 
e-print arXiv:1502.02592 (2015).

\bibitem{ours2}
L. Isaev, Y. H. Moon, and G. Ortiz, Phys. Rev. B \textbf{84}, 075444 (2011); 
M. T. Ahari, G. Ortiz, and B. Seradjeh, e-print arXiv:1508.02682 (2015).

\bibitem{fagotti}
%Local conservation laws in spin-1/2 XY chains with open boundary conditions.
M. Fagotti, e-print arXiv:1601.02011 (2016).

\bibitem{hegde14}
S. Hegde, V. Shivamoggi, S. Vishveshwara, and D. Sen, New J. Phys. {\bf 17}, 053036 (2015).

\bibitem{Viola}
S. Deng, L. Viola, and G. Ortiz, Phys. Rev. Lett. \textbf{108}, 036803 (2012); 
S. Deng, G. Ortiz and L. Viola, Phys. Rev. B \textbf{87}, 205414 (2013).

\bibitem{ours1}
C. W. J. Beenakker, D. I. Pikulin, T. Hyart, H. Schomerus, and J. P. Dahlhaus, 
Phys. Rev. Lett. \textbf{110}, 017003 (2013); 
G. Ortiz, J. Dukelsky, E. Cobanera, C. Esebbag, and C. W. J. Beenakker,
ibid. \textbf{113}, 267002 (2014).

\bibitem{Trench}
% Numerical solution of the eigenvalue problem for Hermitian Toepliz matrices
W. F. Trench, SIAM J. Matrix Anal. Appl. {\bf 10}, 135 (1989).
%D. Bini and V. Pan, Math. Comp. \textbf{50}, 182 (1988).

\bibitem{Ripka}
J.-P. Blaizot and G. Ripka, 
\textit{Quantum Theory of Finite Systems} (MIT Press, Cambridge, MA, 1986).

\bibitem{DFT}
%Tight-Binding Density Functional Theory:  An Approximate Kohn-Sham DFT Scheme
G. Seifert, J. Phys. Chem. A {\bf 111}, 5609 (2007).

\bibitem{Liang}
%Majorana Fermions in Equilibrium and Driven Cold Atom Quantum Wires
L. Jiang, T. Kitagawa, J. Alicea, A. R. Akhmerov, D. Pekker, G. Refael, J. I. Cirac, E. Demler, 
M. D. Lukin, and P. Zoller, Phys. Rev. Lett. {\bf 106}, 220402 (2011).

\bibitem{Symbol}
R. G. Douglas and R. Howe, Trans. Amer. Math. Soc. \textbf{158}, 203 (1971).

\bibitem{next} The Ansatz as presented applies to the generic case where the algebraic 
multiplicity of $z_\ell(\epsilon)$ equals $s_\ell(\epsilon)$, for all $\ell$.  A fully general Ansatz 
along with a rigorous completeness proof of our algorithm will be presented elsewhere.

\bibitem{Non-hermiticity}
J. Bird, R. Kaiser, I. Rotter, and G. Wunner Eds., Special Issue on:
\textit{Quantum Physics with non-Hermitian Operators: Theory and Experiment}, 
Fortschr. Phys. {\bf 61}, 51 (2013). 

\bibitem{note}See Supplemental Material at {[}URL will be inserted
by publisher{]} for additional technical detail on the boundary matrix, 
as well as the BB separation in the Kitaev and $s$-wave chains.

\bibitem{Kitaev01}
A. Kitaev, Phys. Uspekhi \textbf{44}, 131 (2001).

\bibitem{beenakker11}
A. R. Akhmerov, J. P. Dahlhaus, F. Hassler, M. Wimmer and C. W. J. Beenakker, 
Phys. Rev. Lett. {\bf 106}, 057001 (2011).

\bibitem{Remark2}
For the special case $\mu=0$, this solution
was derived in E. Lieb, T. Schultz, and D. Mattis, 
Ann. Phys. \textbf{16}, 407 (1961), using a different technique. 

\bibitem{Flatband}
S. Deng, G. Ortiz, A. Poudel, and L. Viola, Phys. Rev. B {\bf 89}, 140507(R) (2014).

\bibitem{Floquet} 
T. Kitagawa, E. Berg, M. Rudner, and E. Demler, Phys. Rev. B {\bf 82}, 235114 (2010); 
A. Poudel, G. Ortiz, and L. Viola, EPL {\bf 110}, 17004 (2015). 

\bibitem{prosen}
%Third quantization: a general method for solving master equations of quadratic open fermi systems
T. Prosen, New J. Phys. {\bf 10}, 043026 (2008).

\bibitem{ZollerDiss}
S. Diehl, E. Rico, M. A. Baranov, and P. Zoller, Nature Phys. {\bf 7}, 971 (2011).

\end{thebibliography}
\end{document}